\documentclass[preprint,aps,12pt,showpacs,nofootinbib,tightenlines,amsmath,amssymb]{revtex4}

\usepackage{graphicx}
\usepackage{bbm}

\def\be{\begin{equation}}
\def\ee{\end{equation}}
\def\bea{\begin{eqnarray}}
\def\eea{\end{eqnarray}}
\def\bsp{\be\begin{split}}
\def\la{\langle}
\def\ra{\rangle}

\def\bes{\be  \begin{split}}

\def\p{\partial}
\def\la{\langle}
\def\ra{\rangle}
\def\ads{AdS/CFT}
 \makeatletter

\newcommand{\Rmnum}[1]{\expandafter\@slowromancap\romannumeral #1@}
\makeatother

\begin{document}

\title{ Prediction for the Mass Spectra of Resonance Mesons \\ in the Soft-Wall AdS/QCD with a Modified 5D Metric }
\author{Yan-Qin Sui, Yue-Liang Wu, Zhi-Feng Xie and Yi-Bo Yang}
\affiliation{ Kavli Institute for Theoretical Physics China (KITPC)
\\ Key Laboratory of Frontiers in Theoretical Physics
\\ Institute of Theoretical Physics\,\, Chinese Academy of Sciences
\\  Beijing, 100190, People's Republic of China }

\begin{abstract}%%
A soft-wall anti-de Sitter/QCD model with a modified five-dimensional metric at the infrared
region is constructed to obtain a nontrivial dilaton solution,
which incorporates the chiral symmetry breaking and linear
confinement. By taking the pion mass and decay constant as two input
mass scales, the resulting predictions for the resonance states of
pseudoscalar, scalar, vector and axial-vector mesons agree
remarkably with the experimentally confirmed resonance states. The
effects of the quartic interaction term are investigated by taking
an appropriate sign and magnitude for maintaining the stability of
the bulk scalar potential. It is shown that such a simply modified
soft-wall anti-de Sitter/QCD model can lead to a consistent prediction for the
mass spectra of resonance states in the pseudoscalar, scalar, vector
and axial-vector mesons; the agreement with the experimental data is
found to be better than $10\%$ for the excited meson states.
The resulting pion form factor also agrees well with the
experimental data.

\end{abstract}
\pacs{12.40.-y,12.38.Aw,12.38.Lg,14.40.-n}

\maketitle

%%%%%%%%%%%%%%%%%%%%%%%%%%%%%%%%%%%%%%%%%%%%%%%%%%%%%%%%%%%%%%%%%%%%%%

\section{Introduction}

Strong interactions of quarks are described in the standard model
by an $SU(3)$ gauge theory known as quantum chromodynamics
(QCD) \cite{Fritzsch:1973pi}. As the gauge group is non-Abelian, the
gluons have direct self-interactions that lead to the well-known
asymptotic freedom \cite{Gross:1973id,Politzer:1973fx} due to a
negative beta function, $\beta(\mu)$, which causes the coupling
constant $\alpha_{s}(\mu)$ to decrease at short distances (UV
region), so that perturbative QCD at the UV region works well. At
low energies (IR region), perturbative methods are no longer
applicable as the coupling constant $\alpha_{s}(\mu)$ grows in the
IR. We are currently unable to solve from first principle the low
energy dynamics of QCD; one can then construct effective quantum
field theories to describe the low energy features of QCD, such as
dynamically generated spontaneous symmetry breaking
\cite{Nambu:1960xd}. It has been shown in Ref.~\cite{DW} that such a
dynamically generated spontaneous chiral symmetry breaking can lead
to the consistent mass spectra for both the lowest lying nonet
pseudoscalar mesons and nonet scalar mesons. Though the resulting
mass spectra for the ground states were found to agree well with the experimental data, it
is not manifest in a chiral effective field theory how to
characterize the excited meson states.

It is well-known that there are two important features of QCD at the
low energy: they are the chiral symmetry breaking and linear
confinement. It was shown in \cite{'tHooft:1973jz} that for an
$SU(N_c)$ QCD one may carry out a $1/N_c$ expansion in a large $N_c$
limit. In this limit, theory remains maintaining the most important
features such as color confinement and dynamical chiral symmetry
breaking. Thus any consistent low energy QCD model should
simultaneously characterize these two basic features. Besides that,
the model should also practically be applicable to calculate the low
energy quantities of QCD, such as the decay and coupling constants,
the mass spectra of various meson resonances.

The duality between gravity and gauge theories conjectured by
Maldacena \cite{Maldacena:1997re} and further developed in
\cite{Gubser:1998bc,Witten:1998qj} has shed new light on solving the
problem of strongly coupled gauge theories. Thus the anti-de
Sitter/conformal field theory (AdS/CFT) conjecture is regarded as an
important step in theoretical physics in the past ten years,
which establishes the duality between the weak coupled supergravity
in $AdS_5$ and the strong coupled ${\cal N}=4$ super Yang-Mills, and
thus makes the calculations in the strong coupled theory become
feasible \cite{Polchinski:2001tt}. This feature has attracted a lot of
attention recently. It is expected that an analogous duality holds
between AdS and QCD, though the latter is not an exact conformal
theory. There are two approaches to pursuing this duality: one is the
so-called top-down approach \cite{Sakai:2004cn} and the other is the
bottom-up approach \cite{Erlich:2005qh,DP}. The former starts with
the string theory and varies the gravity background so as to
reproduce the basic QCD features. The latter is inspired by the
AdS/CFT conjecture and known as a phenomenological AdS/QCD model.
The model consists of a gauge theory in a curved space (usually AdS)
with the field contents chosen to holographically match some
bound states and operators in QCD.  It is also interesting to
observe the correspondence between
matrix elements obtained in AdS/CFT with the corresponding formula
using the light-front representation as shown in Refs. \cite{BT1,BT2,BT3}.

The current impressive achievements of AdS/QCD models contain the
chiral symmetry breaking in a hard-wall AdS/QCD model
\cite{Erlich:2005qh} and the linear confinement in a soft-wall
AdS/QCD model \cite{Karch:2006pv}. Nevertheless, in the hard-wall
model \cite{Erlich:2005qh}, the resulting mass spectra for the
excited mesons are contrary to the experimental data. In the
soft-wall model \cite{Karch:2006pv}, one can obtain a desired mass
spectra for the excited vector mesons, while the chiral symmetry
breaking phenomenon cannot consistently be realized. Note that a
dilaton field in the soft-wall model is introduced by hand as a
uniform background field. Interesting progress was made in
Refs.~\cite{Colangelo:2008us,Gherghetta:2009ac}: a quartic
interaction term in the bulk scalar potential was introduced to
incorporate linear trajectories and chiral symmetry breaking.
Nevertheless, such a term was shown~\cite{Gherghetta:2009ac} to
cause an instability of the scalar potential and result in a
negative mass for the lowest lying scalar meson state and much
smaller mass spectra for other lowest lying meson states in
comparison with the experimental data. Thus how to naturally
incorporate these two important features into a single AdS/QCD model
and obtain the consistent mass spectra remains a challenging and
interesting task.

On this note, we provide an alternative soft-wall AdS/QCD model by
simply modifying the five-dimensional (5D) metric in the IR region. The
paper is organized as follows: In Sec. \ref{sec:2}, we introduce
the modified 5D soft-wall AdS/QCD model and show how the background
dilaton field gets a desired IR behavior from a simply modified 5D
metric; several phenomenological AdS/QCD models corresponding to
different choices of the bulk vacuum expectation value (VEV) of the
scalar field are considered. In Sec. III, we provide a detailed
analysis and show how such simply modified soft-wall AdS/QCD models
can simultaneously describe both chiral symmetry breaking and linear
confinement, and lead to a reasonable prediction for the mass
spectra of the various resonance states in the pseudoscalar, scalar,
vector and axial-vector mesons. Unlike the predictions given in
\cite{Gherghetta:2009ac}, the simply modified AdS/QCD model in our
present considerations contains no virtual meson state in the scalar
sector. In particular, the resulting resonance meson states agree
well with the experimentally confirmed meson states and there are
no more additional unconfirmed resonance meson states existing in
the present model. The effects of a quartic interaction term in the
bulk scalar potential are investigated in Sec. IV. By taking an
appropriate sign (that is opposite to the one considered in
\cite{Gherghetta:2009ac}) for the coupling constant of the quartic
interaction to keep the stability of the bulk scalar potential, we
found that the resulting predictions for the mass spectra of
resonance mesons can be further improved and consistent with
the experimental data. Our conclusions and remarks are presented in
the last section.

\section{The Soft-Wall AdS/QCD Model with Modified 5D
Metric}\label{sec:2}

In the real world, QCD is known to be neither supersymmetry nor
quantum mechanically conformal. Therefore, the 5D space of AdS in
the AdS/QCD model is not necessary to be a pure AdS. Here we shall
consider the simplest extension to a 5D AdS with the following
metric structure
 \be \label{metric}
ds^2=a^2(z)\left(\eta_{\mu\nu}dx^\mu dx^\nu-dz^2\right);\qquad
a^2(z)=(1+\mu_g^2\,z^2)/z^2
 \ee
where $\eta_{\mu\nu}=\rm{diag}\left(1,-1,-1,-1\right)$, and $\mu_g$
is a constant mass scale. Such a nonpure AdS space was also
considered in \cite{Shock:2006qy} with a hard-wall cut. Here we
shall show that in the soft-wall AdS/QCD with the above simply
modified 5D metric at the IR region, it can lead to a consistent
prediction for the mass spectra in the pseudoscalar, scalar, vector
and axial-vector meson resonances. The effects of a quartic term in
the bulk scalar potential are found to further improve the mass
spectra of resonance mesons when taking an appropriate sign and
magnitude for the coupling constant.

It has been shown in \cite{Karch:2006pv} that the introduction of a
background dilaton $\Phi$ can lead to a linear trajectory for
resonance vector meson mass once $\Phi$ has an asymptotic behavior
\be\label{dilation0}
 \Phi(z\to\infty)=\mu_d^2 \,z^2
 \ee
where the parameter $\mu_d$ sets the meson mass scale. The 5D action
with the background field of dilaton $\Phi(z)$ and a quartic term in
the bulk scalar potential can be written as follows
 \be\label{lagran}
S_5=\int d^5x\sqrt{g}e^{-\Phi(z)}\,{\rm{Tr}}\left[|DX|^2-m_X^2|X|^2-
\lambda |X|^4 - \frac1{4g_5^2}\left(F_L^2+F_R^2\right)  \right]
 \ee
with $g=|\det g_{MN}|$, $D^MX=\p^MX-i A_L^MX+i X A_R^M$,
$A_{L,R}^M=A_{L,R}^{M~a}t^a$ and ${\rm{Tr}}[t^at^b]=\delta^{ab}/2$. Here
$A_{L,R}^{M}$ are introduced to gauge the chiral symmetry
$SU(2)_L\times SU(2)_R$. $\lambda$ is the coupling constant which has an opposite sign in comparison with the one in
\cite{Gherghetta:2009ac}.

The parameter $g_5$ is fixed to be $g_5^2 = 12\pi^2/N_c$
\cite{Erlich:2005qh} with $N_c$ the color number and $m_X^2=-3$ by
\ads ~ correspondence. The UV boundary condition for the gauge
fields $A_L$ and $A_R$ at $z=0$ is given by the value of the sources
of the currents $J_L$ and $J_R$ in 4D theory as required by the
holographic correspondence. The IR boundary condition with the
background field dilaton playing the role of the smooth soft-wall
cutoff simply requires that the action is finite at $z\to \infty$,
so that the ambiguity of the choice of the IR boundary condition in
a hard-wall theory \cite{Erlich:2005qh} disappears in the soft-wall
theory \cite{Karch:2006pv}.

The VEV of $X$ field in 5D space has the following form for the two
flavor case
    \begin{displaymath}
     \la X(z) \ra=\frac{1}{2}\,v(z)\left( \begin{array}{cc}
                      1 & 0 \\
                      0 & 1 \\
                   \end{array} \right).
    \end{displaymath}
For simplicity, we shall first consider the case with $\lambda =0$.
In this case, the VEV $v(z)$ satisfies the following condition
 \be\label{xexpect}
 \p_z\left(a^3(z)e^{-\Phi}\p_zv(z)
 \right)-a^5(z)e^{-\Phi}m_X^2v(z)=0.
 \ee
For the role played by $X$ field on the boundary of $AdS_5$
\cite{Klebanov:1999tb}, the VEV $v(z)$  has the following behavior
at the UV boundary $z\to 0$:
 \be\label{vzero}
 v(z\to0)=m_q \,\zeta\, z+\frac{\sigma\, z^3}{\zeta}\\
 \ee
where $m_q$ and $\sigma$ are interpreted by AdS/CFT duality as the
quark mass and quark condensate respectively. The normalization
$\zeta$ is fixed by QCD with $\zeta=\sqrt{3}/(2\pi)$
\cite{Damgaard:2008zs}. The corresponding solution for the dilaton
field at the UV boundary is found from Eq.\,(\ref{xexpect}) to be

\bes\label{dilaton}
 \Phi'(z\to 0)&= 6\mu_g^{2}\,z + O(z^3),\qquad \Phi(z\to 0) = 3\mu_g^{2}\,z^2 + O(z^4)
 \end{split} \ee

The behavior of the VEV $v(z)$ in the IR boundary $z\to\infty$ is
correlated to the dilaton behavior given in Eq.~(\ref{dilation0})
which affects the mass spectra of meson resonances. In our present
consideration with a modified metric given in Eq.~(\ref{metric}), it
is not difficult to find from Eq.~(\ref{xexpect}) that a polynomial
leading behavior for $v(z\to\infty)$ is enough to result in the
required IR boundary condition for the dilaton background field. In
general, we have the following IR boundary condition for $v(z)$
 \bes\label{vinfty}
 v(z\to\infty)&= \gamma\, (\mu_d z)^{\alpha},
 \end{split} \ee
with $\alpha$ being a positive parameter. Instituting the above
boundary condition into Eq.\,(\ref{xexpect}), we can obtain a
solution for the dilaton at the IR boundary:
 \bes\label{dilaton2}
 \Phi'(z\to\infty)&=\frac{3\mu_g^{2}}{\alpha}\,z, \qquad \mu_d^2 =
 \frac{3}{2} \mu_g^2/\alpha
 \end{split} \ee

\begin{figure}[ht]
\begin{center}
\includegraphics[width=10cm,clip=true,keepaspectratio=true]{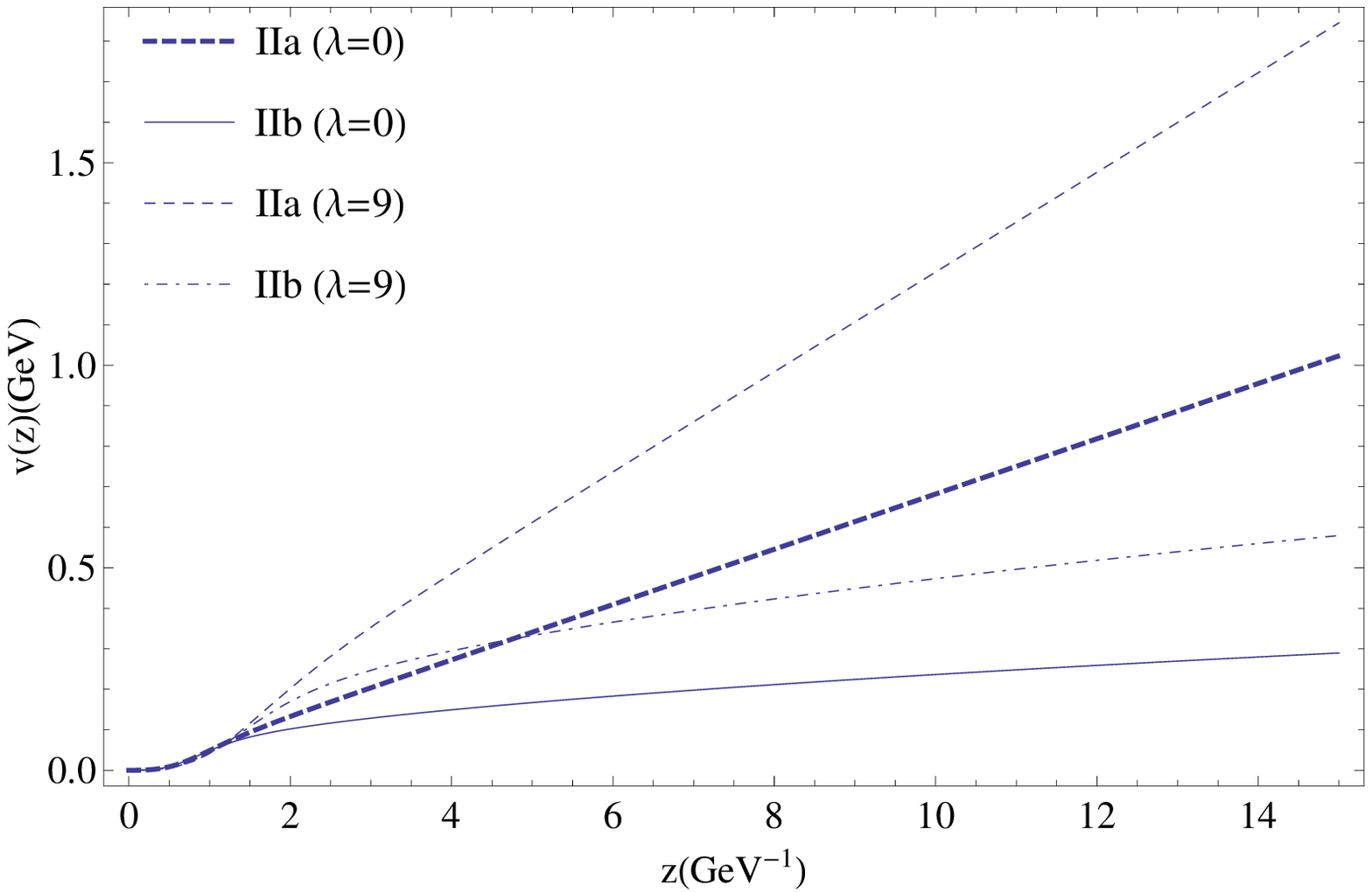}
\caption{A plot of $v(z)$ for various parameters which is fitted to
the mass spectra.}\label{fig:gvz}
\end{center}
\end{figure}

\begin{figure}[ht]
\begin{center}
\includegraphics[width=10cm,clip=true,keepaspectratio=true]{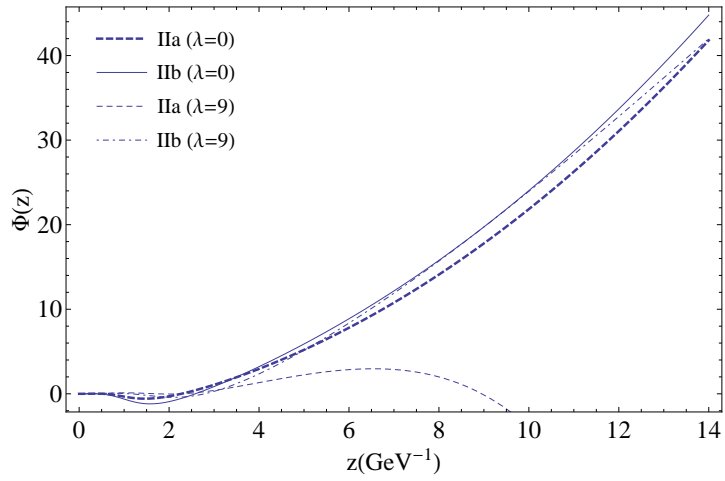}
\caption{A plot of dilation $\Phi(z)$ for various parameters fitted
to the mass spectra  }\label{fig:gdilaton}
\end{center}
\end{figure}

To investigate the dependence of the mass spectra of resonance
mesons on the IR boundary conditions of the VEV $v(z)$, we are going
to consider two interesting asymptotic behaviors of $v(z)$ at the IR
boundary in three typical models which correspond to three
different exact forms (I, II, III) of the VEV $v(z)$. One
corresponds to $\alpha=1$ as shown in the models Ia, IIa, IIIa,
and the other to $\alpha=\frac{1}{2}$ as shown in the models Ib,
IIb, IIIb. Explicitly, two different asymptotic behaviors of $v(z)$
are given by
 \bes\label{cases}
 \mbox{Case a}: \quad v(z\to\infty)= \gamma\, (\mu_d z);
 \qquad \mbox{Case b}:\quad   v(z\to\infty)= \gamma\, (\sqrt{\mu_d
 z}).
 \end{split} \ee
The explicit forms of $v(z)$ for three types of models (I, II, III)
with two IR boundary conditions are summarized in the Table\,
\ref{formsofv}.
\begin{table}[ht!]
\begin{center}
\begin{tabular}{ccc}
\hline\hline
       Models & $v(z)$          & $\qquad$ Parameters \\
\hline
       Ia& $ z(A+Bz^2)(1+Cz^2)^{-1} $&$B=\frac{\sigma }{\zeta}+m_q\zeta C$,$C=B/\mu_d\gamma$\\

       Ib& $ z(A+Bz^2)(1+Cz^2)^{-5/4}$ & $B=\frac{\sigma }{\zeta}+\frac{5}{4}m_q\zeta C$,$C=(B^2/\mu_d\gamma^2)^{2/5}$ \\

       IIa&$ z(A+Bz^2)(1+Cz^4)^{-1/2}$& $B=\frac{\sigma }{\zeta}$, $C=(B/\mu_d\gamma)^2$ \\

       IIb&$ z(A+Bz^2)(1+Cz^4)^{-5/8}$& $B=\frac{\sigma }{\zeta}$,$C=(B^2/\mu_d\gamma^2)^{4/5}$ \\

      IIIa&$z[A+B \tanh(Cz^2)]$&  $B=\mu_d\gamma-m_q\zeta$,$C=\frac{\sigma }{\zeta B}$\\

      IIIb &$z[A+B \tanh(Cz^2)](1+Gz^4)^{-1/8}$&  $B=\mu_d^{1/2} \gamma G^{1/8}-m_q\zeta$,$C=\frac{\sigma }{\zeta B}$\\
\hline\hline
\end{tabular}
\caption{Three type of models for $v(z)$ with two cases for each
type of model and relevant parameters with $A=m_q\zeta$ for all the
cases.} \label{formsofv}
\end{center}
\end{table}
The three quantities $m_q$, $\sigma$ and $\gamma$ appearing in the
boundary conditions of $v(z)$ are mainly correlated to the three
parameters $A$, $B$ and  $C$ ( and $G$ in IIIb case). The model III
was shown to be a well parametrized one in the modified soft-wall
model \cite{Gherghetta:2009ac}. It will be shown below that the
results in our present considerations are not very sensitive to the
exact forms of the bulk VEV $v(z)$; they mainly depend on the IR
boundary conditions. For a comparison, we plot in Fig.~\ref{fig:gvz}
the bulk VEV $v(z)$ for two cases in model II. The corresponding
dilaton field is plotted in Fig.~\ref{fig:gdilaton} as the function
of $z$. The three parameters $m_q$, $\sigma$ and $\gamma$ (or $A$,
$B$ and $C$) are fixed by the known experimental values of $m_\pi
=139.6$ MeV and $f_\pi=92.4$ MeV with minimizing the breaking of
the Gell-Mann-Oakes-Renner relation $f_{\pi}^2m_{\pi}^2 = 2m_q
\sigma$ at the $1\%$ level. The parameter $G$ is obtained by
optimizing the mass spectra of vector and axial-vector mesons in
model IIIb.

The pion decay constant is calculated from the axial-vector equation
of motion with the pole in the propagator set to zero which was
discussed in detail in Ref.~\cite{Erlich:2005qh}. Note that the axial-vector equation of motion depends on both quark mass and condensate.
The pion decay constant is given \cite{Erlich:2005qh}:
\begin{equation}
f_\pi^2=\left.-\frac{1}{g_5^2}\frac{\partial_z
A(0,z)}{z}\right|_{z\to 0}\ ,
\end{equation}
where $A(0,z)$ is the axial-vector bulk-to-boundary propagator and is obtained by solving the equation of motion in the momentum space
 \be
 e^{\Phi}\partial_z\left(a(z)e^{-\Phi}\partial_z A(q,z)\right) + a(z) q^2 A(q,z)
-a^3(z)g_5^2 v^2(z)A(q,z)=0
 \ee
with $q^2=0$ and the boundary conditions $A(0,0)=1$ and
$\p_zA(0,z\to\infty)=0$. The pion mass is related to the pseudoscalar equation of motion for
the lowest lying state and will be discussed in Sec. \ref{sec:pi}.

\section{Mass Spectra of Pseudoscalar, Scalar, Vector and Axial-Vector Mesons}\label{sec:eom}

In this section, we are going to make numerical calculations for the
mass spectra of pseudoscalar, scalar, vector and axial-vector
mesons. As $\mu_d$ or $\mu_g$ scales the mass spectra of meson
resonances, it is not difficult to find out its value from a global
fitting, the best value for the case $\lambda=0$ is found to be
\bes \mu_d = 445\, \rm{MeV}; \quad \mu_g = 363\, \rm{MeV} \, \quad
\mbox{(Case a)},\qquad \mu_g = 257\, \rm{MeV} \, \quad \mbox{(Case
b)}
 \end{split} \ee
The values of three fitting parameters $m_q$, $\sigma$ and $\gamma$
are presented in the Table\ \ref{parameter}. The mass spectra for
the pseudoscalar, scalar, vector and axial-vector resonance mesons
are given in Tables III, IV, V and VI. All the quoted experimental
data are taken from the particle data group (PDG)
\cite{Amsler:2008zzb}. It is seen that the resonance states agree
well with the experimentally confirmed states and the resulting mass
spectra are consistent with the experimental values, except for the
ground states of the scalar and axial-vector mesons which have
masses smaller than the experimental data. The effects of a quartic
interaction term in the bulk scalar potential are going to be
studied in the next section and shown to be necessary for further
improving the mass spectra.
\begin{table}[ht!]
\begin{center}
\begin{tabular}{ccccccc}
\hline\hline
Parameter &Ia&Ib& IIa & IIb  & IIIa & IIIb  \\
\hline 
         $m_q$  (MeV)     &   4.16         &   4.64         &       4.44 &       4.07 &       4.98 &     4.25 \\

$\sigma^{\frac{1}{3}}$ (MeV) &    275        &   265         &        265 &        272 &        255 &   268 \\

$\gamma$              &    0.178     &  0.136         &  0.153      &   0.112    &  0.164     & 0.112  \\
\hline\hline
\end{tabular}
\caption{The parameters $m_q$,  $\sigma$ and $\gamma$ for
$\lambda=0$ in three type of models with two cases $a$ and $b$
corresponding to the IR boundary conditions given in Eq.
(\ref{cases}). }\label{parameter}
\end{center}
\end{table}
\subsection{Pseudoscalar Mesons}\label{sec:pi}

Writing the bulk scalar field as
$X(x,z)\equiv(v(z)/2+S(x,z))e^{2i\pi(x,z)}$ with $S(x,z)$ being the
scalar meson field and $\pi(x,z)=\pi^a(x,z) t^a$ being the
pseudoscalar meson field, and decomposing the axial field in terms
of its transverse and longitudinal components, $A_{\mu}^{a}=A_{\mu
\bot}^{a}+\partial_{\mu}\phi^{a} $,  we then obtain, from the action
Eq.\,(\ref{lagran}), the following equation of motion in the 4D
momentum space with the $A_5=0$ gauge:
\bes\label{pi1}
&\partial_z\left(a(z)e^{-\Phi}\partial_z\phi^{a}\right)+g_5^2\,a^3(z)\,v^2(z)e^{-\Phi}\,(\pi^{a}-\phi^{a})=0\\
& q^2\partial_z\phi^{a}-g_5^2\,a^2(z)\,v^2(z)\partial_z\pi^{a}=0
\end{split}\ee
By eliminating the longitudinal component field $\phi$ from the
above coupled equation Eq.~(\ref{pi1}), we obtain the following
equation for the $\pi$ field
 \be \label{pit}
 -\partial_z^2 \tilde{\pi}(q,z)+ V_{\pi}(z)~\tilde{\pi}(q,z)= q^2\tilde{\pi}(q,z),
 \ee
 where $\tilde{\pi}(q,z)\equiv \partial_z \pi(q,z)$ and
 \be  \label{piv}
 \begin{split}
 V_{\pi}(z)&=g_5^2
 a^2(z)v^2(z)+\frac{\Phi'^2+2\Phi''}{4}+\frac{15a^{'2}(z)}{4a^2(z)}-\frac{3a'(z)(v(z)\Phi'-2v'(z))}{2a(z)v(z)}-\frac{3a''(z)}{2a(z)}\\
 &+\frac{2v^{'2}(z)}{v^2(z)}-\frac{\Phi'v'(z)+v''(z)}{v(z)}.
 \end{split}
 \ee
Assuming $\tilde{\pi}(q,z)=\sum_{n} \Pi_n(q) \tilde{\pi}_n(z)$, we
arrive at the following equation of motion
 \be \label{pin}
 -\partial_z^2 \tilde{\pi}_n(z)+ V_{\pi}(z)~\tilde{\pi}_n(z)= m_{\pi_n}^2 \tilde{\pi}_n(z),
 \ee
where $q^2$ is replaced by $m_{\pi_n}^2$ with $m_{\pi_n}$ being
the masses of pseudoscalar mesons.  The above equation can be solved
by the shooting method. Using the boundary conditions
$\tilde{\pi}(z\to 0) =0$, $\partial_z \tilde{\pi}(z\to \infty) =0$,
we obtain the mass spectra of excited states with the input of $\pi$
mass and decay constant. The numerical results are given in
Table.~\ref{pseudoscalarmasses} and also plotted in
Fig.~\ref{fig:gmass1}.
 \begin{table}[ht!]
\begin{center}
\begin{tabular}{cccccccc}
\hline\hline
         n & $\pi$~experimental.~(MeV) & Ia &Ib & IIa &IIb  &IIIa &IIIb  \\
\hline 
        0 &      139.6 &      139.6 &      139.6 &      139.6 &      139.6 &      139.6 &         139.6 \\

        1 &$1350 \pm 100$&       1219 &       1127 &       1285 &       1474 &       1339 &          1524 \\

        2 &$1816 \pm 14$ &       1632 &       1483 &       1664 &       1733 &       1721 &        1779 \\

       3&     ------         &      1949      &    1754        &   1960         &  1958        &    2015          &    1999         \\

        4&    ------            &    2218        &     1983       &   2212         &   2169       &     2264         &    2207         \\

       5&     ------           &       2455     &      2185      &    2444        &    2388      &      2491        &     2422        \\

       6&     ------           &     2670       &      2368      &    2677        &    2627      &    2717          &     2652        \\

       7&     ------           &     2869       &      2536      &     2923       &    2881      &      2956        &    2897         \\
\hline\hline
\end{tabular}
\caption{The experimental and predicted mass spectra for
pseudoscalar mesons with $\lambda=0$.} \label{pseudoscalarmasses}
\end{center}
\end{table}
Where the mass of pion meson is as an input, the resulting excited
meson states agree well with the data in models IIb, IIIa and IIIb.
The agreement is seen to be within $10\%$; it is then interesting to
provide a prediction for possible high excited states.

\subsection{Scalar Mesons}
 % eom

Assuming $X(x,z)\equiv(v(z)/2+S(x,z))e^{2i\pi(x,z)}$ and
$S(x,z)=\sum_{n}\mathcal {S}_n(x)S_n(z)$, we arrive at the following
equation of motion
  \be
  \p_z\left(a^3(z)e^{-\Phi}\p_z S_n(z)
  \right)-a^5(z)e^{-\Phi}m_X^2S_n(z)=-a^3(z)e^{-\Phi}m_{S_n}^2S_n(z)
  \ee
By defining $ S_n(z)\equiv e^{\omega_s/2}s_n(z)=e^{(\Phi-3\log a(z)
)/2}s_n(z)$, we have
   \be\label{scalareom}
  -\p_z^2s_n(z)+\left(\frac1{4}\omega_s'^2-\frac1{2}\omega_s''+a^2(z)m_X^2
  \right)s_n(z)=m_{S_n}^2s_n(z)
   \ee
For simplicity, we consider here only the mass spectra for the
$SU(3)$ singlet scalar mesons which have more experimental data.

\begin{figure}[hb]
\begin{center}
\includegraphics[width=11cm,clip=true,keepaspectratio=true]{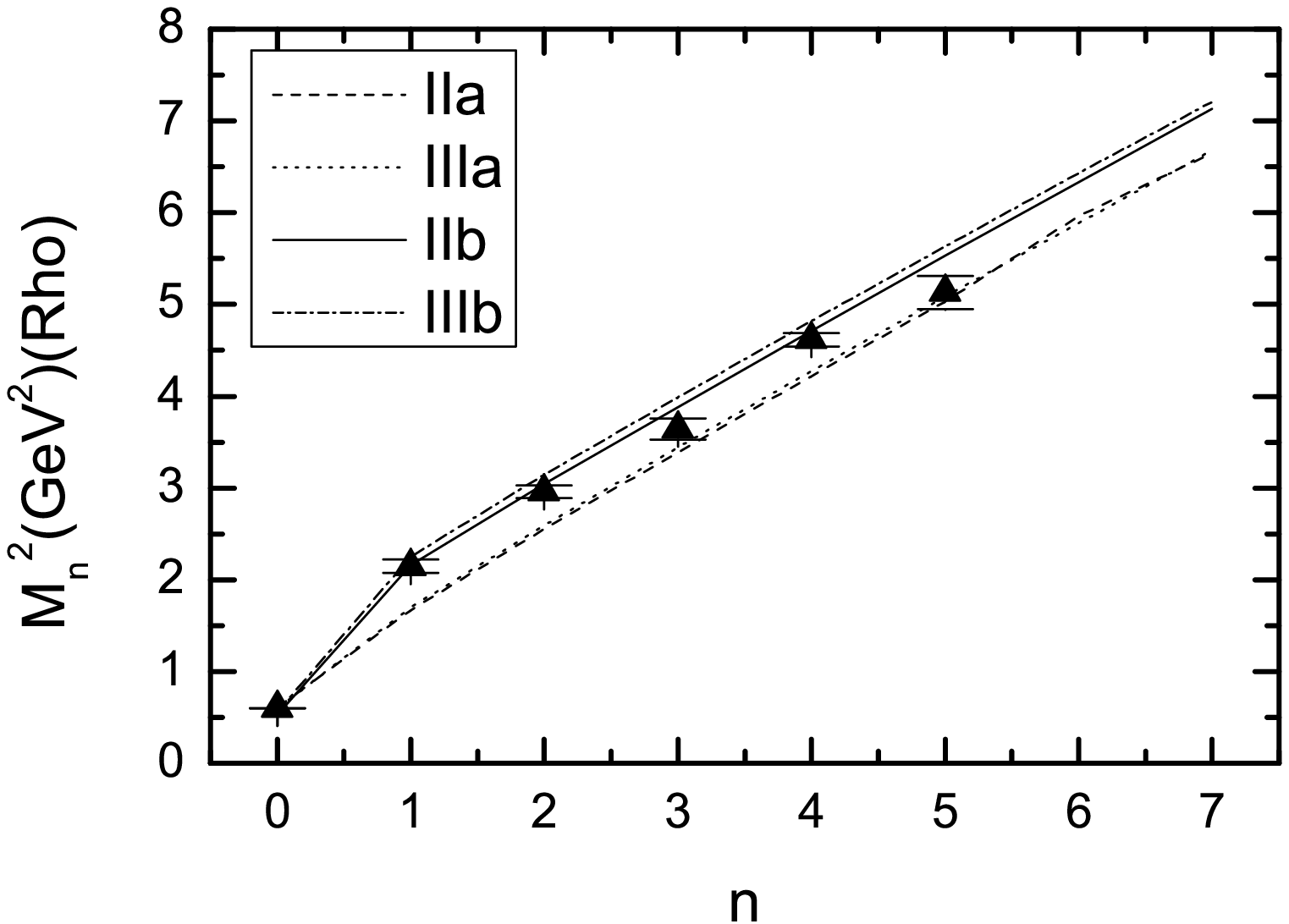}
\includegraphics[width=11cm,clip=true,keepaspectratio=true]{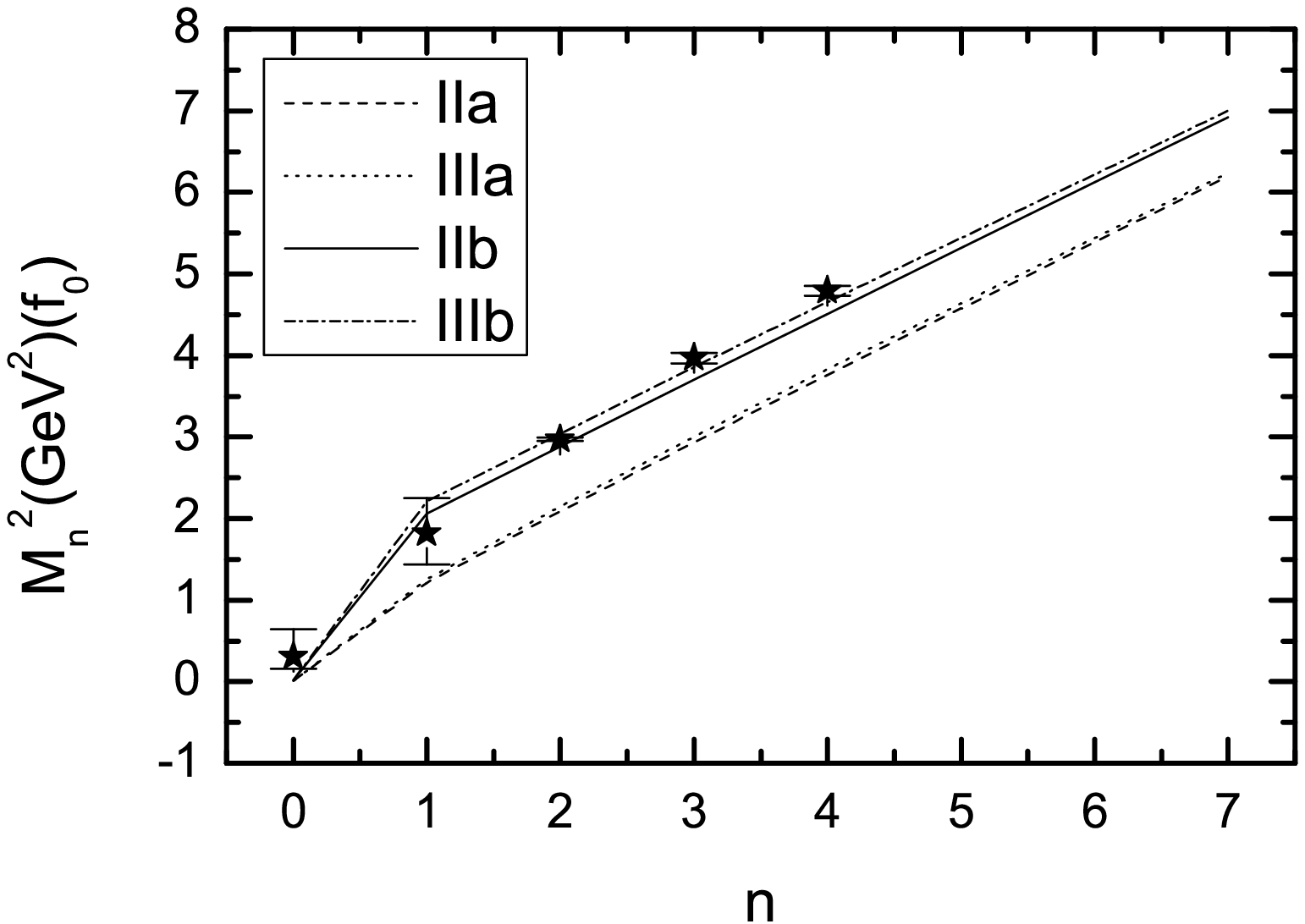}
\includegraphics[width=11cm,clip=true,keepaspectratio=true]{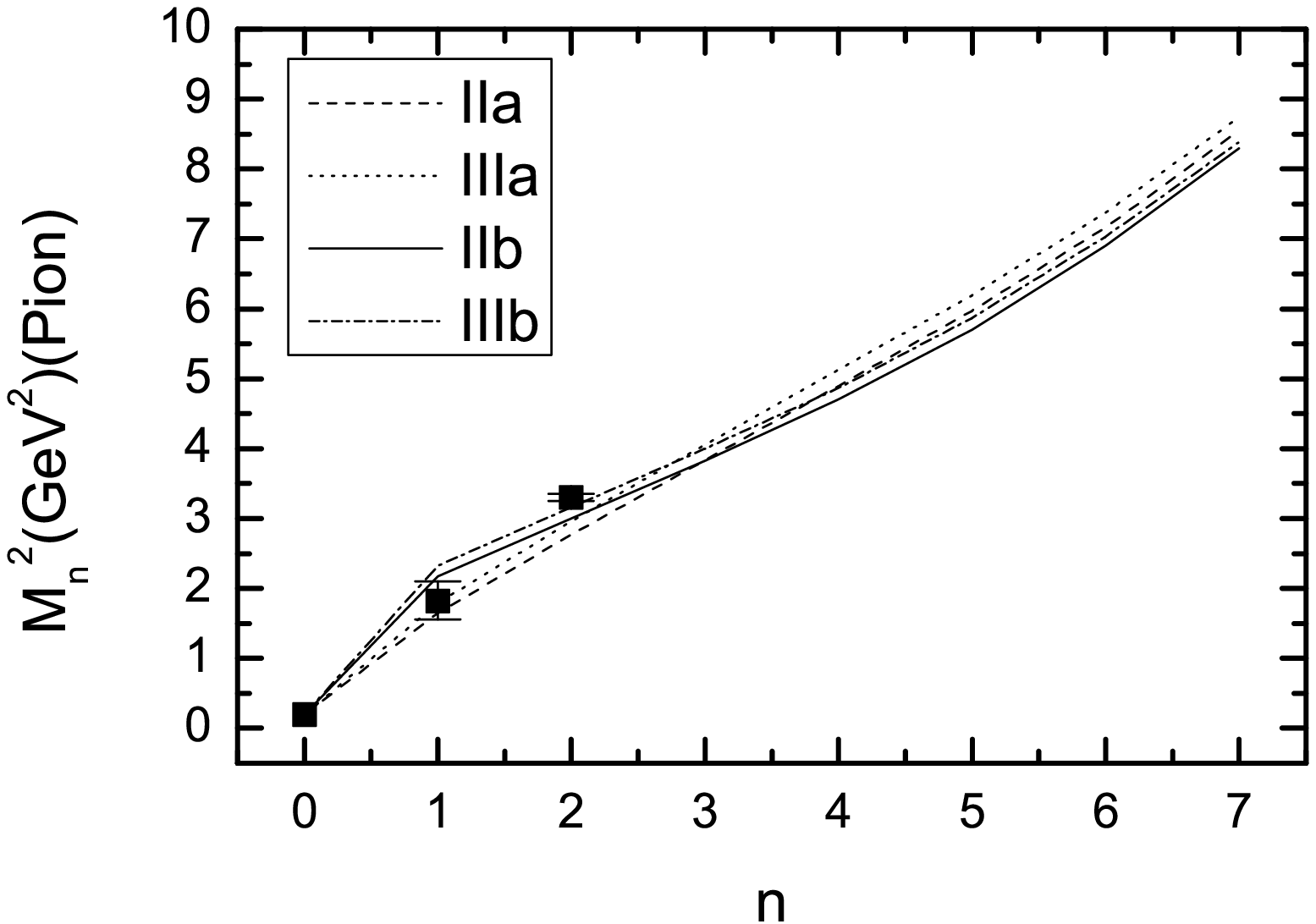}
\includegraphics[width=11cm,clip=true,keepaspectratio=true]{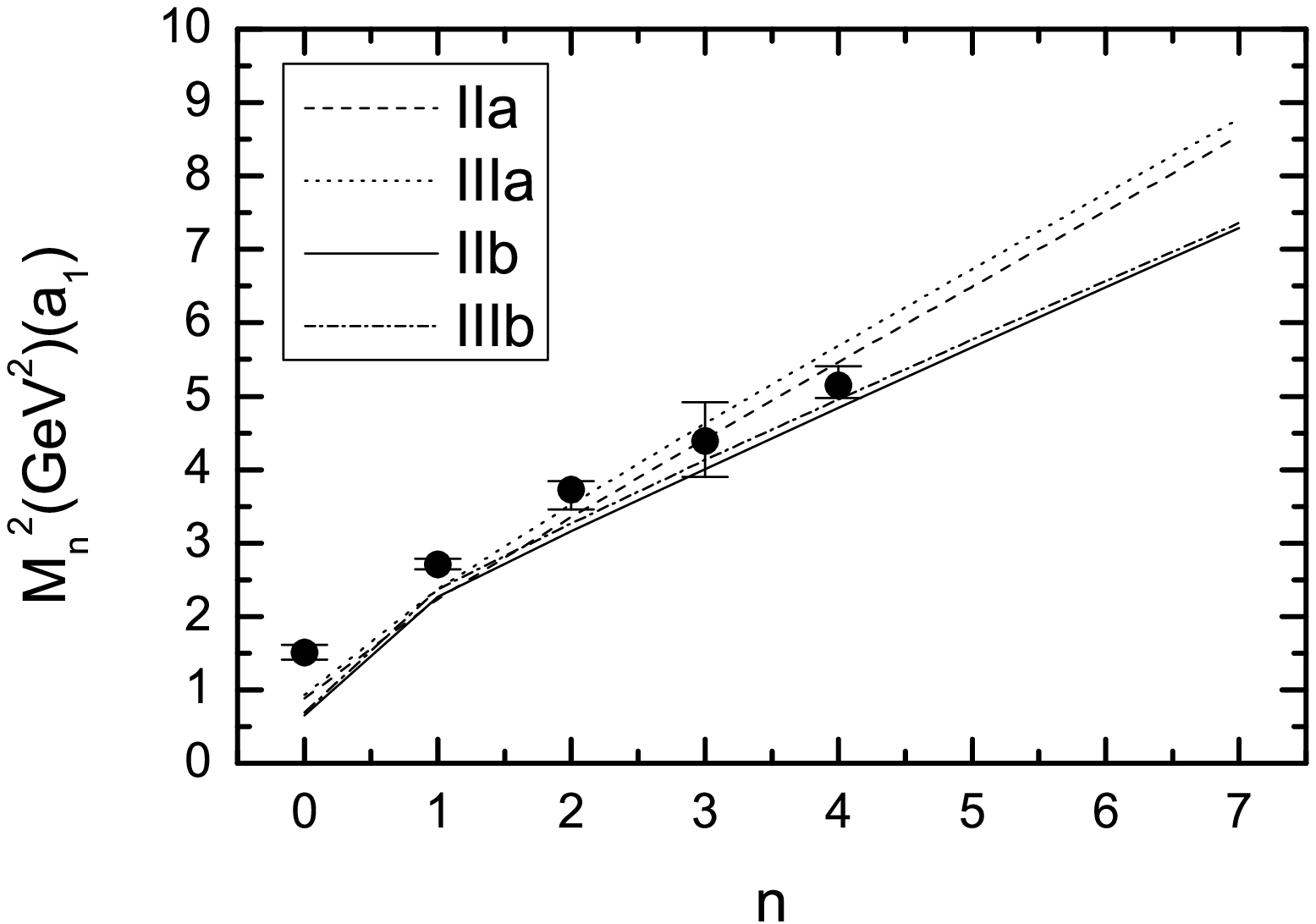}
\caption{A plot of mass spectra of resonance mesons in models
IIa$(\lambda=0)$, IIIa$(\lambda=0)$, IIb$(\lambda=0)$,
IIIb$(\lambda=0)$.  The squares are the experimental mass spectra
for the resonance pseudoscalars ($\pi$), the triangles for the
resonance vector mesons ($\rho$), the circles for the resonance
axial-vector mesons ($a_1$), and the stars are for the resonance scalar
mesons ($f_0$). }\label{fig:gmass1}
\end{center}
\end{figure}

Using the shooting method to solve Eq.\,(\ref{scalareom}) with the
boundary conditions $s_n(z\to 0) =0$, $\partial_z s_n(z\to \infty)
=0$, we obtain the mass spectra for the $SU(3)$ singlet resonance
scalar mesons which are given in Table \ref{scalarmasses} and also
plotted in Fig. \ref{fig:gmass1}.
%
% mass eigenvalue
\begin{table}[ht!]
\begin{center}
\begin{tabular}{cccccccc}
\hline\hline
        n & $f_0$~experimental.~(MeV)& Ia &Ib & IIa &IIb  &IIIa &IIIb  \\
\hline
       0& $550^{+250}_{-150}$  &      115 &        121 &        119 &        127 &        115 &         126 \\

         1 & $1350 \pm 150$     &       1002&       1050 &       1099 &       1434 &       1122 &         1485 \\

         2 &  $1724 \pm 7$                  &       1366 &       1418 &       1446 &       1697 &       1468 &       1743 \\

         3 &  $1992 \pm 16$        &       1644 &       1693 &       1713 &       1923 &       1734 &        1964 \\

         4 & $2189 \pm 13$       &       1877 &       1925 &       1939 &       2124 &       1958 &          2158 \\

       5&     ------           &     2083       &     2128       &    2140        &   2306       &     2155         &   2333          \\

      6&      ------          &      2268      &      2312      &    2321        &     2474     &      2333        &      2494       \\

      7&      ------          &    2440        &      2482      &   2489         &    2630      &       2497       &     2645        \\
\hline\hline
\end{tabular}
\caption{The experimental and predicted mass spectra for the singlet
scalar mesons with $\lambda=0$.} \label{scalarmasses}
\end{center}
\end{table}

\begin{figure}[ht]
\begin{center}
\includegraphics[width=8cm,clip=true,keepaspectratio=true]{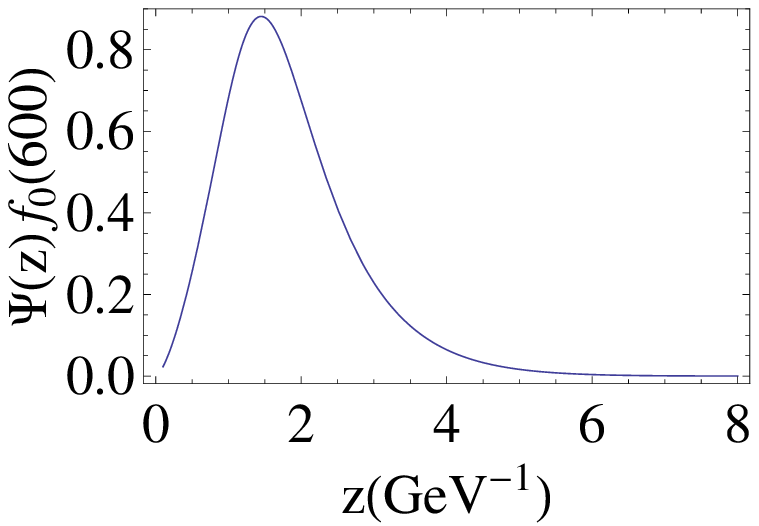}
\includegraphics[width=8cm,clip=true,keepaspectratio=true]{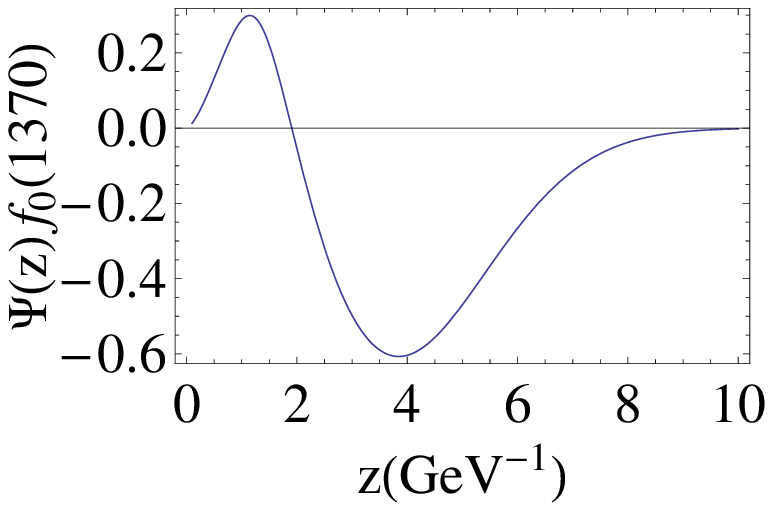}
\includegraphics[width=8cm,clip=true,keepaspectratio=true]{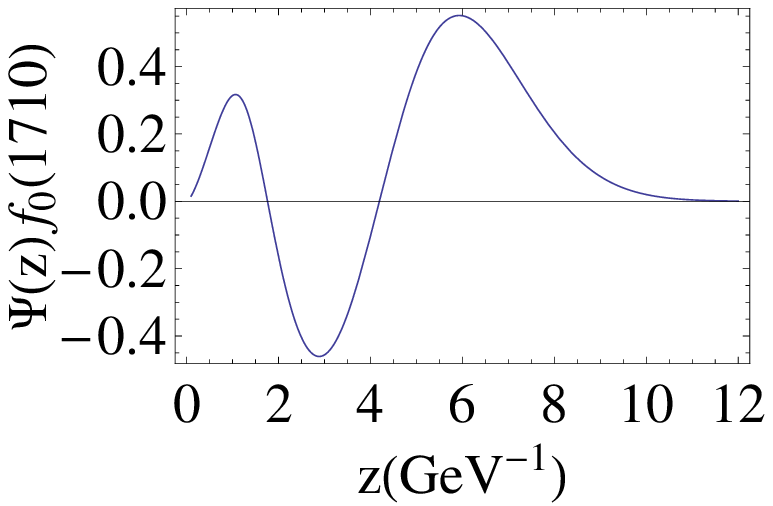}
\includegraphics[width=8cm,clip=true,keepaspectratio=true]{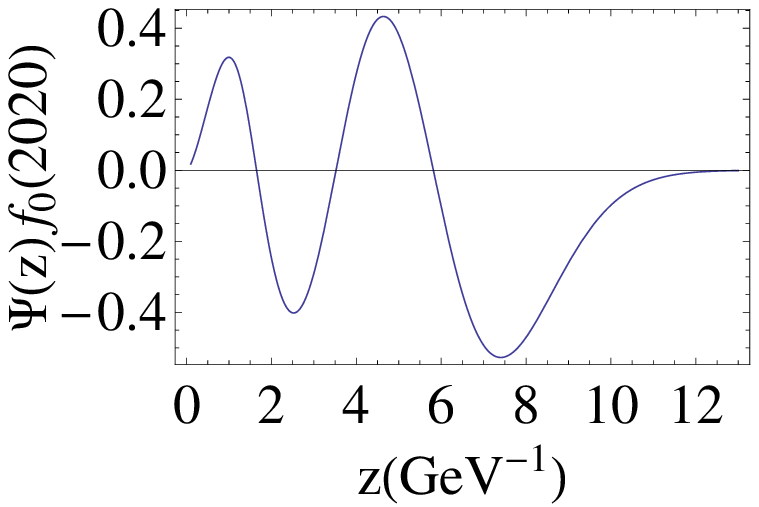}
\caption{Normalized bulk wave function for scalar mesons in case
IIb$(\lambda=0)$. }\label{fig:gwavef}
\end{center}
\end{figure}

It is seen that the resonance states agree well with the
experimentally confirmed states though the ground state has a small
mass (about 120 MeV) in comparison with the experimental data
$550^{+250}_{-150}$ MeV, which have the biggest uncertainties. Note
that this is unlike the model only considering a quartic interaction in
the bulk scalar potential \cite{Gherghetta:2009ac}, where it was
shown that the model may cause an instability and contains virtual
mass for the lowest lying scalar meson. This is avoided in the
present modified models which contain no virtual mass state. It must
be clarified that the scalar states $f_0 (980 \pm 10)$, $f_0(1505
\pm 6 )$, $f_0(2103 \pm 8 )$ and $f_0(2314 \pm 25 )$ should be
classified into the isosinglet resonance scalar states of $SU(3)$
octet mesons, rather than the $SU(3)$ singlet resonance scalar
states. As an interesting check, we plot in Fig.~\ref{fig:gwavef}
the corresponding bulk wave functions of $SU(3)$ singlet resonance
scalar mesons; it is seen that the oscillation property becomes
manifest in the scalar sector. Note that the possible instanton
effects and the mixing effects between the $SU(3)$ singlet scalar
and the isosinglet scalar of the $SU(3)$ octet are not included in the
present considerations. It is manifest from Fig.~\ref{fig:gmass1}
that the models IIb and IIIb lead to a better agreement with the
experimental data.

\subsection{Vector Mesons}
\begin{figure}[ht]
\begin{center}
\includegraphics[width=10cm,clip=true,keepaspectratio=true]{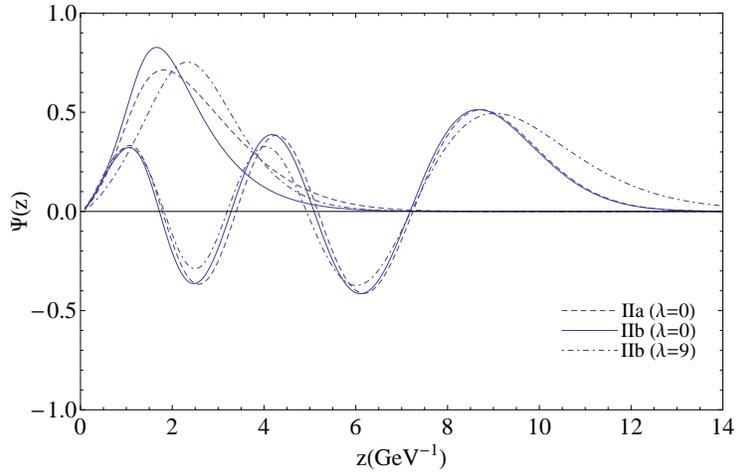}
\caption{Normalized bulk wave function for vector mesons with the
single peak curve for the bulk wave function of ground state $\rho$
and multipeaks for the excited vector meson corresponding to $n=4$.
}\label{fig:gwavev}
\end{center}
\end{figure}

 % eom
  From the action Eq.\,(\ref{lagran}), with the gauge fixing $V_5=0$, one can
  derive the equation of motion for vector field
  \be\label{veom}
  -\p_z^2V_n+\omega'\p_zV_n=m^2_{V_n}V_n,
  \ee
For simplicity,  we omit the Lorentz index and group index in flavor
space. Defining $V_n\equiv e^{\omega/2} v_n=e^{\left(\Phi(z)-\log
a(z)\right)/2}v_n$, the above equation can be rewritten as
 \be
 -\p_z^2v_n+\left(\frac1{4}\omega'^2-\frac1{2}\omega''\right)v_n=m^2_{V_n}v_n.
 \ee
Such an eigenvalue equation can also be solved by the shooting
method, using the boundary conditions $v_n(z\to 0) =0$, $\partial_z
v_n(z\to \infty) =0$; the resulting mass spectra are presented in
Table.\,\ref{vectormasses} and also plotted in
Fig.~\ref{fig:gmass1}. It is interesting to note that such a simple
model can lead to a remarkable agreement with the experimental data,
especially in the models IIb and IIIb. For an illustration, we also
plot in Fig.~\ref{fig:gwavev} the bulk wave functions of the ground
state and the excited state $(n=4)$ for various cases.
\begin{table}[ht!]
\begin{center}
\begin{tabular}{cccccccc}
\hline\hline
        n & $\rho$~experimental.~(MeV) & Ia &Ib  & IIa &IIb  &IIIa &IIIb  \\
\hline
         0 &  $775.5 \pm 1$  &        739 &        603 &        777 &        727 &        775 &            748 \\

         1 &  $1465 \pm 25$ &       1223 &       1175 &       1292 &       1468 &       1303 &           1501 \\

         2 & $1720 \pm 20$  &       1534 &       1509 &       1596 &       1744 &       1610 &          1773 \\

         3 & $1909 \pm 30$   &       1784 &       1769 &       1842 &       1971 &       1856 &         1999 \\

         4 & $2149 \pm 17$  &       2000 &       1990 &       2054 &       2170 &       2068 &        2196 \\

         5 & $2265 \pm 40$  &       2193 &       2187 &       2249 &       2351 &       2255 &        2373 \\

     6&      ------          &   2370         &   2367         &   2417         &   2516       &  2426            &   2535          \\

    7&       ------         &   2534         &    2532        &      2578      &    2671      &    2584          &    2685         \\
\hline\hline
\end{tabular}
\caption{The experimental and predicted mass spectra for vector
mesons with $\lambda =0$.} \label{vectormasses}
\end{center}
\end{table}

\subsection{Axial-vector Mesons}
 % eom

From the action Eq.\,(\ref{lagran}) with the gauge $A_5=0$, one can
derive the equation of motion for perpendicular component of axial
field
\be\label{Atransverse}
e^{\Phi}\partial_z(a(z)e^{-\Phi}\partial_zA_{n})+a(z)q^2A_{n}-a^3(z)g_5^2v^2(z)A_{n}=0
\ee
Again defining $A_n\equiv e^{\omega/2} a_n=e^{\left(\Phi(z)-\log
a(z)\right)/2}a_n$, the above equation of motion can be reexpressed
as
  \be\label{axialeom}
 -\p_z^2a_n+\left(\frac1{4}\omega'^2-\frac1{2}\omega''+g_5^2v^2(z)a^2(z)\right)a_n=m^2_{A_n}a_n.
  \ee
With the boundary conditions $a_n(z\to 0) =0$, $\partial_z a_n(z\to
\infty) =0$, the resulting mass spectra by using the shooting method
is given in Table.\,\ref{avmasses} and also plotted in Fig.
\ref{fig:gmass1}. The resonance states agree well with the
experimental ones, while the mass for the ground state is slightly
smaller than the experimental data.

\begin{table}[ht!]
\begin{center}
\begin{tabular}{cccccccc}
\hline\hline
         n & $a_1$~experimental.~(MeV) & Ia &Ib  & IIa &IIb  &IIIa &IIIb  \\
\hline
         0 & $1230 \pm 40$            &       934 &        714 &        940 &        807 &        963 &         833 \\

         1 & $1647 \pm 22$           &       1468 &       1247 &       1496 &       1507 &       1539 &        1540 \\

         2 & $1930^{+30}_{-70}$           &       1822 &       1573 &       1831 &       1778 &       1880 &       1807 \\

         3 &   $2096 \pm 122$           &      2109 &       1829 &       2102 &       2003 &       2152 &        2031 \\

         4 &  $2270^{+55}_{-40}$          &      2358 &       2049 &       2338 &       2202 &       2386 &       2228 \\

        5&     ------        &  2582          &   2247         &    2549        &  2380        &   2594           &   2403          \\

       6&      ------          &    2787        &    2439        &       2742     &   2545       &      2785        &     2564        \\

        7&     ------          &    2979        &    2638        &      2922      &  2699        &     2964         &    2713         \\
\hline\hline
\end{tabular}
\caption{The experimental and predicted mass spectra for
axial-vectors with $\lambda=0$.} \label{avmasses}
\end{center}
\end{table}

\newpage

\section{Quartic Interaction of Bulk Scalar}

%%%%%%%%%%%%%%%%%%%%%%%%%%%%%%%%%%%%%%%%%%%%%%%%%%%%%%%%%%%%%%%%%%%%%%

%%%%%%%%%%%%%%%%%%%%%%%%%%%%%%%%%%%%%%%%%%%%%%%%%%%%%%%%%%%%%%%%%%%%%%

It is interesting to note that the above simplest AdS/QCD model with
four parameters can lead to a consistent prediction for all the
experimentally confirmed resonance meson states, while the ground
state masses of scalar and axial-vector mesons obtained above appear
to be smaller than the experiment data though the lowest lying
scalar mass has the biggest uncertainty. To make a possible
improvement, we now turn to consider the effects of the quartic
interaction $\lambda |X|^4$ in the bulk scalar potential. In this
case, the equation for the VEV $v(z)$ is modified to be
 \be\label{xexpect1}
 \p_z\left(a^3(z)e^{-\Phi}\p_zv(z)
 \right)-a^5(z)e^{-\Phi}(m_X^2+\frac{\lambda}{2} v^2(z))v(z)=0.
 \ee
and the equation of motion for the scalar field is extended to be
\be
  \p_z\left(a^3(z)e^{-\Phi}\p_z S_n(z)
  \right)-a^5(z)e^{-\Phi}\left(m_X^2+\frac{3}{2}\lambda v^2(z)\right) S_n(z)=-a^3(z)e^{-\Phi}m_{S_n}^2S_n(z)
  \ee

\begin{figure}[ht]
\begin{center}
\includegraphics[width=11cm,clip=true,keepaspectratio=true]{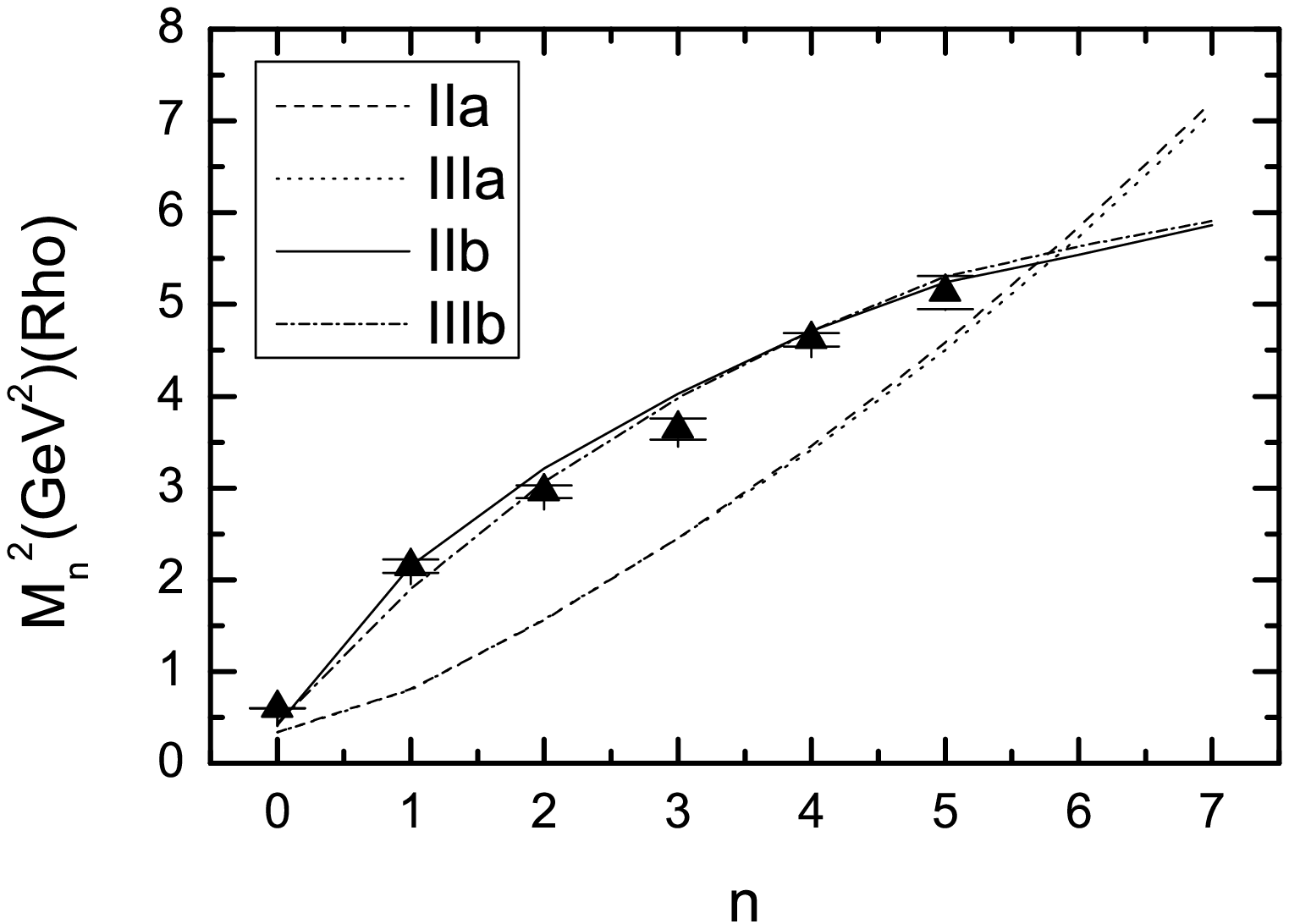}
\includegraphics[width=11cm,clip=true,keepaspectratio=true]{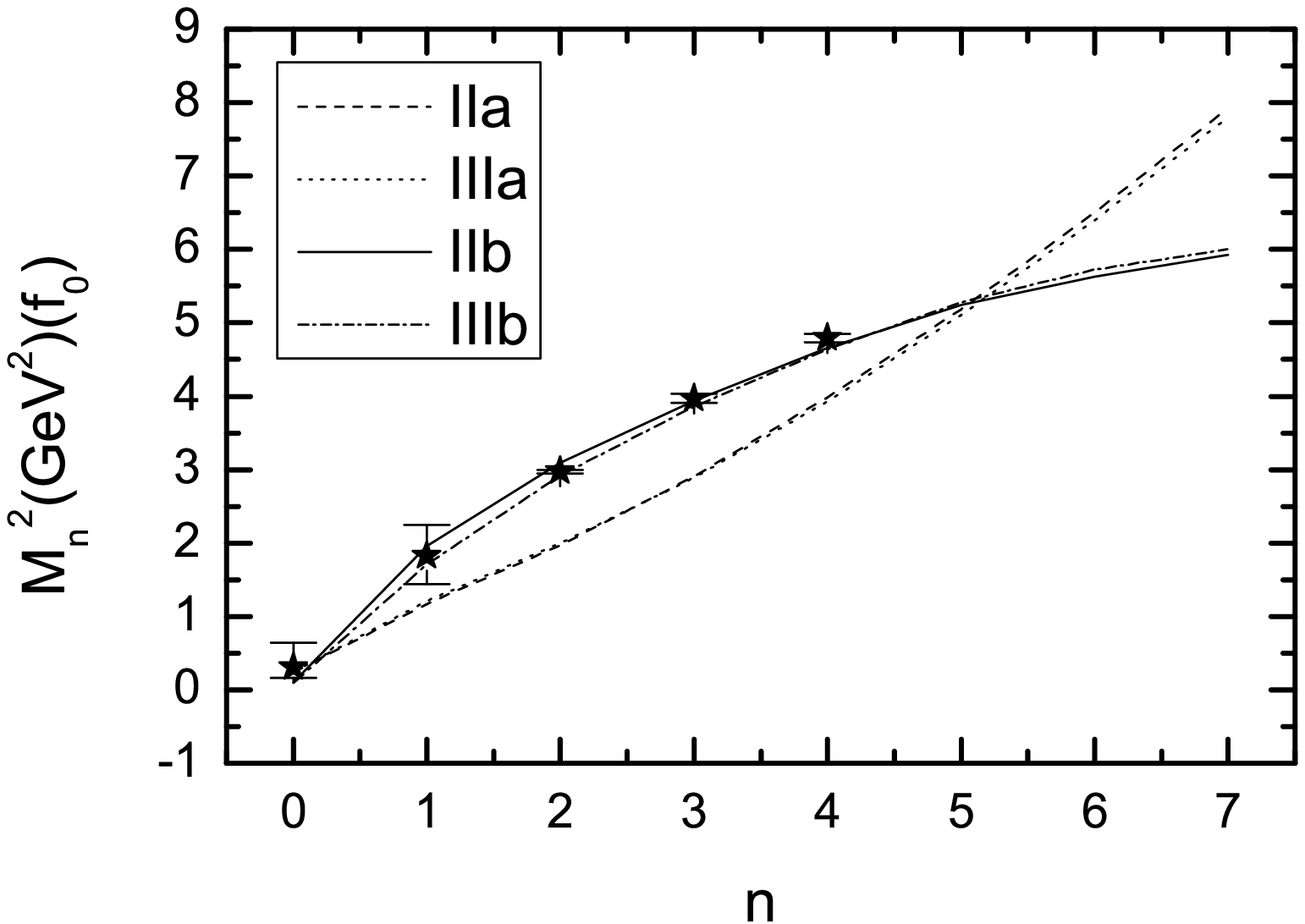}
\includegraphics[width=11cm,clip=true,keepaspectratio=true]{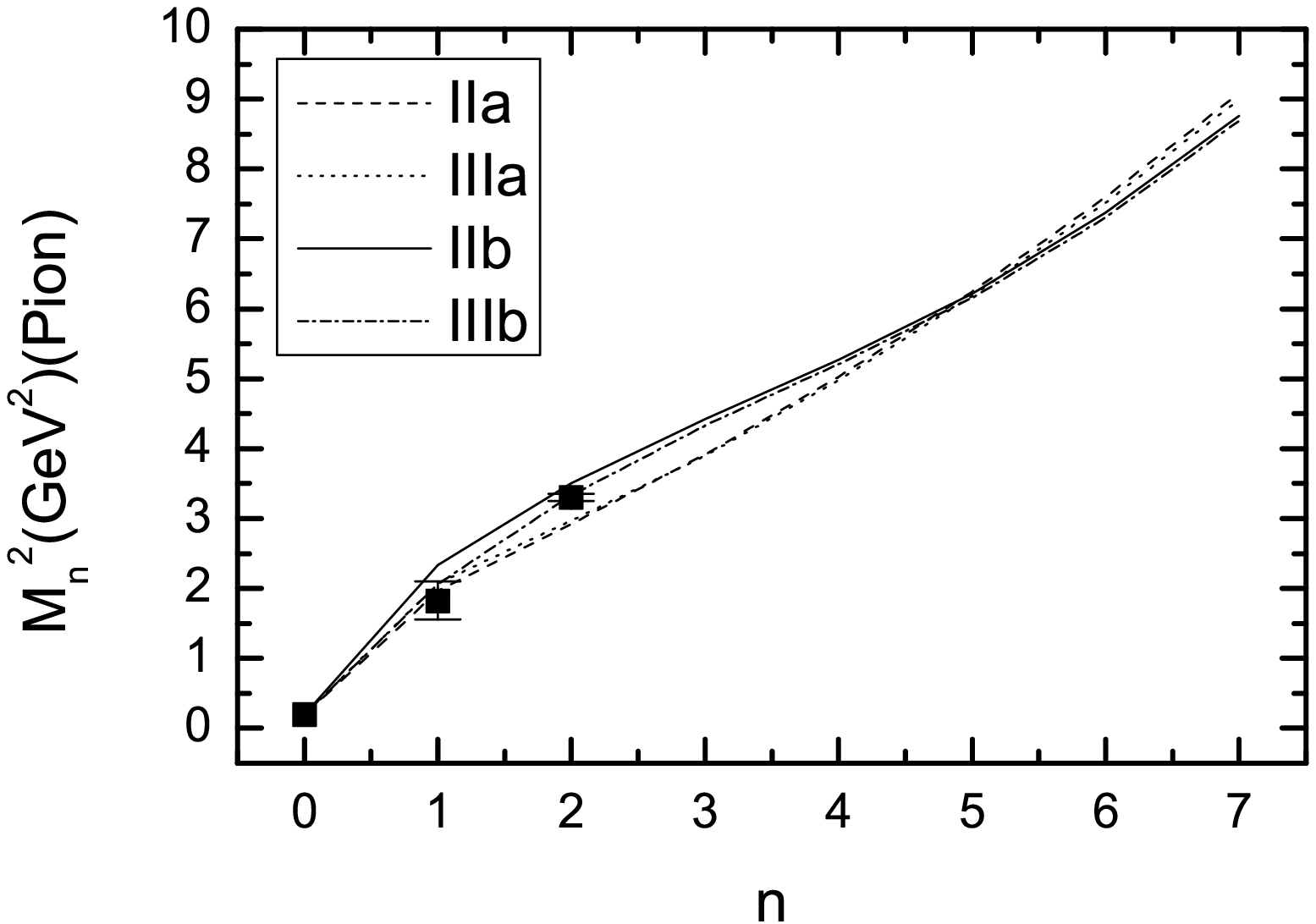}
\includegraphics[width=11cm,clip=true,keepaspectratio=true]{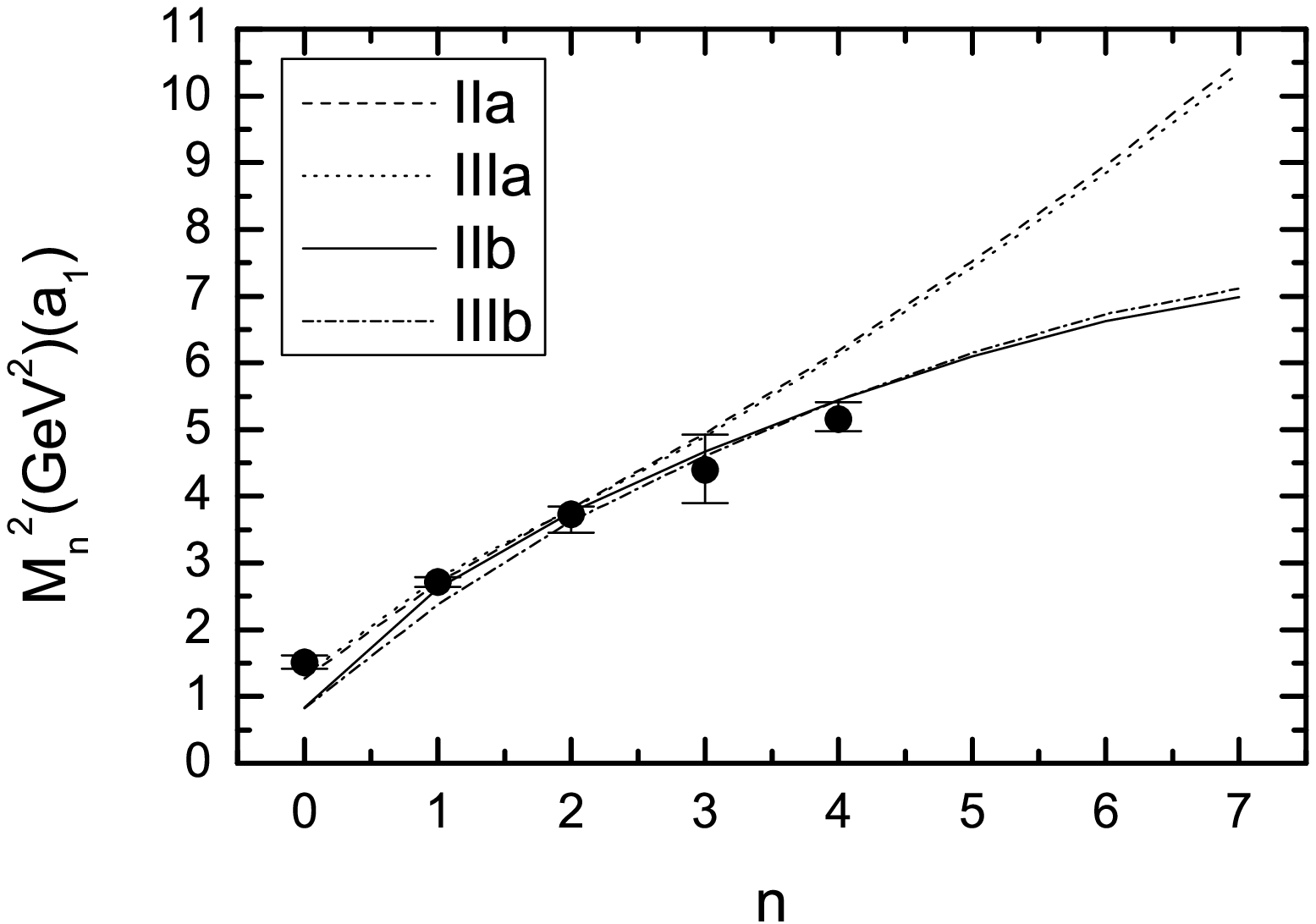}
\caption{Similar to Fig.~\ref{fig:gmass1}, a plot of mass spectra of
resonance mesons in models IIa$(\lambda=9)$, IIIa$(\lambda=9)$,
IIb$(\lambda=9)$, IIIb$(\lambda=9)$.  }\label{fig:gmass2}
\end{center}
\end{figure}

Using the shooting method and making a global fitting with input
mass scales of the $\pi$ meson mass and decay constant as well as the Gell-Mann-Oakes-Renner
relation, we present all the numerical results in Tables VIII, IX,
X, XI and also plot them in Fig.~\ref{fig:gmass2}.  The model
parameters are reanalyzed and given in Table VII. The reasonable
value for the coupling constant $\lambda$ is found to be $\lambda
=9$ and the parameter $G$ in model IIIb is fitted to be
$G=0.01~\rm{GeV}^4$.

\begin{table}[ht!]
\begin{center}
\begin{tabular}{ccccc}
\hline\hline
Parameter &  IIa & IIb  &  IIIa & IIIb \\
\hline
           $m_q$ (MeV)               &       6.95 &       6.79 &      7.08 &       6.49 \\

           $\sigma^{\frac{1}{3}}$ (MeV)&       228 &        229 &        226 &        233  \\

           $\gamma$              &       0.30 &       0.20  &      0.29 &       0.20 \\

           $\mu_d$ (MeV)         &       412 &        548 &       412  &       557 \\
\hline\hline
\end{tabular}
\caption{The fitting parameters $m_q$,  $\sigma$,  $\gamma$, $\mu_d$
with $\lambda =9$. } \label{parameterk}
\end{center}
\end{table}

It is interesting to see that the inclusion of the quartic
interaction term with an appropriate sign and magnitude can further
improve the predictions. Especially, the models IIb and IIIb lead to
a better agreement, which shows that the IR boundary condition
$v(z\to \infty) \sim \sqrt{z}$ is more reasonable than the IR
boundary condition $v(z\to \infty) \sim z$.

\begin{table}[ht!]
\begin{center}
\begin{tabular}{cccccc}
\hline\hline
     n &$\pi$\, experimental.~(MeV)& IIa  & IIb   & IIIa & IIIb  \\
\hline
         0 &      139.6 &               139.6 &      139.6 &      139.6 &      139.6 \\

         1 &      $1350 \pm 100$ &       1399 &       1526 &       1441 &       1437  \\

         2 &       $1816 \pm 14$ &       1709 &       1871&      1722 &       1824 \\

       3&     ------         &      1979     &    2103      &  1975        &  2080      \\

        4&    ------            &    2242       &    2295      &   2231        &   2283     \\

       5&     ------           &      2500     &     2493     &   2488       &    2482   \\

       6&     ------           &    2757       &    2716     &  2743       &    2704    \\

       7&     ------           &    3013       &      2960     &   2998    &    2948     \\
\hline\hline
\end{tabular}
\caption{The experimental and predicted mass spectra for
pseudoscalar mesons with $\lambda=9$.} \label{pseudoscalarmassesk}
\end{center}
\end{table}

\begin{table}[ht!]
\begin{center}
\begin{tabular}{cccccc}
\hline\hline
         n & $f_0$\, experimental.~(MeV)  & IIa &IIb  &IIIa &IIIb \\
\hline
         0 & $550^{+250}_{-150}$  &  493     & 317   &  495    & 305   \\

         1 & $1350 \pm 150$     &  1083       &  1401  &  1100    & 1307   \\

         2 &  $1724 \pm 7$                &    1402     &   1757 &   1413   &  1708  \\

         3 &  $1992 \pm 16$    &     1707    &  1986  &   1702   &  1964  \\

         4 & $2189 \pm 13$    &    1997    &  2158  &  1982    &  2154  \\

       5&     ------           &      2278     &     2290     &   2260       &    2297  \\

       6&     ------           &    2550       &    2373     &  2531      &    2392    \\

       7&     ------           &    2815      &      2434    &   2793   &    2448     \\
\hline\hline
\end{tabular}
\caption{The experimental and predicted mass spectra for scalar
mesons with $\lambda=9$.} \label{scalarmassesk}
\end{center}
\end{table}

\begin{table}[ht!]
\begin{center}
\begin{tabular}{cccccc}
\hline\hline
         n &$\rho$\, experimental.~(MeV) & IIa & IIb  & IIIa & IIIb \\
\hline
         0 &  $775.5 \pm 1$  & 583     &  646    & 584     & 661   \\

         1 &  $1465 \pm 25$ &  900     &   1468   & 893     & 1378   \\

         2 & $1720 \pm 20$  &   1248    &   1793   &  1252    & 1753   \\

         3 & $1909 \pm 30$   &   1564    &   2008   &  1565    & 1994  \\

         4 & $2149 \pm 17$  &   1860    &    2170   &  1847    &  2174  \\

         5 & $2265 \pm 40$  &   2143    &     2289  &   2122   &  2332  \\

       6&     ------           &   2417      &   2353   &  2395     &   2372   \\

       7&     ------           &   2685     &    2420    &  2661   &  2432    \\
\hline\hline
\end{tabular}
\caption{The experimental and predicted mass spectra for vector
mesons with $\lambda=9$. } \label{vectormassesk}
\end{center}
\end{table}

\begin{table}[ht!]
\begin{center}
\begin{tabular}{cccccc}
\hline\hline
         n &$a_1$\, experimental(MeV)&  IIa  & IIb  & IIIa & IIIb \\
\hline
         0 & $1230 \pm 40$       &  1128      & 913   & 1150     & 906   \\

         1 & $1647 \pm 22$      &   1643     &  1618  &  1668    & 1544   \\

         2 & $1930^{+30}_{-70}$  &   1953    &  1940  &  1950    &  1903  \\

         3 &   $2096 \pm 122$    &   2225    &   2161 &   2213   &  2146  \\

         4 &  $2270^{+55}_{-40}$   &  2486   &   2333 &   2473   &  2332  \\

       5&     ------           &      2742    &     2470     &   2726       &    2480  \\

       6&     ------           &    2993      &    2574     &  2974      &    2594    \\

       7&     ------           &    3239      &      2642    &   3217   &    2667     \\
\hline\hline
\end{tabular}
\caption{The experimental and predicted mass spectra for
axial-vector mesons with $\lambda=9$.} \label{avmassesk}
\end{center}
\end{table}

%\newpage

\section{Vector Coupling and Pion Form Factor}

As a useful check for the consistency of our present AdS/QCD model,
we are going to perform a calculation for the vector coupling
$g_{\rho\pi\pi}$ and the pion form factor $F_{\pi}(q^2)$. The vector
coupling $g_{\rho\pi\pi}$ in the soft-wall AdS/QCD is given as
follows\cite{Erlich:2005qh}:
 \be\label{grhopipi}
 g_{\rho\pi\pi}=\frac{g_{5}}{N}\int dz\,
 V_{\rho}(z)\,e^{-\Phi(z)}\left(\frac{a(z)(\partial_z\varphi(z))^2}{g_{5}^2}+v^2(z)a^3(z)(\pi(z)-\varphi(z))^2\right)
 \ee
where $V_{\rho}(z)$ is the $\rho$ meson bulk wave function
corresponding to the vector meson bulk wave functions $V_{n}$ with
$n=0$. Note that in all the integrals over $z$, the integration region is in
principle the whole range $z\in (0,~\infty)$. While in the practical calculations, for the lower limit, one can set a value as small as possible so as to obtain a stable result, for the upper limit, considering the suppressed factor
$e^{-\Phi(z)}$, one can take at a certain finite value where the integrand tends to be zero. $V_n$ satisfies the
normalization condition:
    \be
   \int dz \,a(z)\,e^{-\Phi(z)}V_{n}(z)V_{m}(z)=\delta_{mn}.
    \ee
The functions $\pi(z)$ and $\varphi(z)$ are the solutions of
Eq.\,(\ref{pi1}) and normalized as follows:
 \be
  N= \int dz\,
e^{-\Phi(z)}\,\left(\frac{a(z)(\partial_z\varphi(z))^2}{g_{5}^2}+v^2(z)a^3(z)(\pi(z)-\varphi(z))^2\right)
\ee
The numerical results are found to be $g_{\rho\pi\pi}= 3.63$ MeV
(IIa),  $2.94$ MeV (IIb), $3.81$ MeV (IIIa), $3.12$ MeV (IIIb) with
$\lambda=0$. After including the quartic interaction with
$\lambda=9$, the results are changed to be $g_{\rho\pi\pi} = 2.86$
MeV (IIa), $3.51$ MeV (IIb), $2.93$ MeV (IIIa), $3.53$ MeV (IIIb),
which are still below the experimental value
$g_{\rho\pi\pi}=6.03\pm 0.07$ MeV. Other soft-wall models also
obtained small values, for instance, $g_{\rho\pi\pi} = 2.89$ in
\cite{Gherghetta:2009ac}.

\begin{figure}[hd]
\begin{center}
\includegraphics[width=12cm,clip=true,keepaspectratio=true]{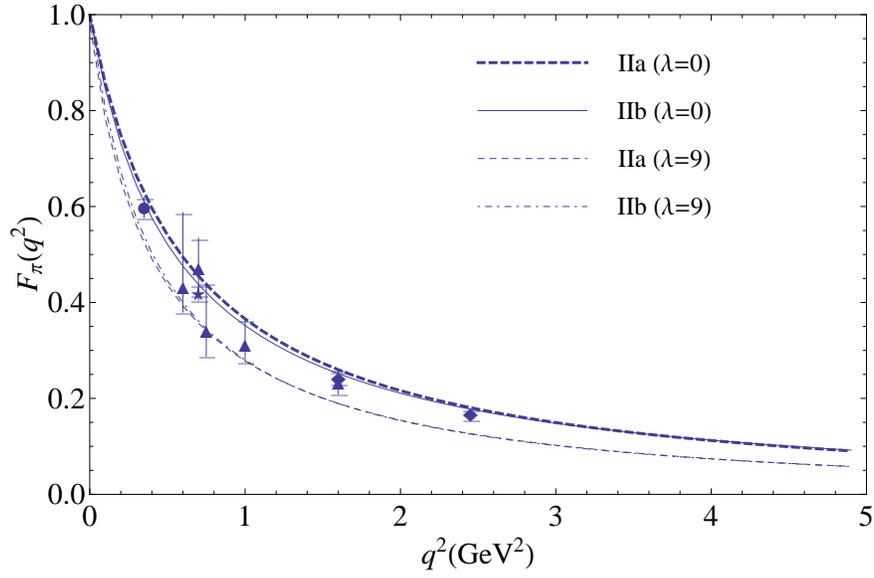}
\caption{The predicted spacelike behavior of the pion form factor
$F_\pi(q^2)$ compared to the experimental data analyzed in
\cite{Kwee:2007nq}. The triangles are data from DESY, reanalyzed by
\cite{Tadevosyan:2007yd}. The diamonds are data from Jefferson Lab
\cite{Horn:2006tm}.  The circle \cite{Ackermann:1977rp} as well as
 the star \cite{Brauel:1977ra} are also data obtained from DESY.}\label{fig:gfpi}
\end{center}
\end{figure}

We now carry out a calculation for the spacelike pion form factor
$F_\pi(q^2)$ by adopting the expressions in terms of the vector and
axial-vector bulk-to-boundary propagators as given in
\cite{Kwee:2007nq}:
\be\label{Fpi}
 F_\pi(q^2)=\frac{g_{5}}{N}\int dz\,
 V(q,z)\,e^{-\Phi(z)}\left(\frac{a(z)(\partial_z\varphi(z))^2}{g_{5}^2}+v^2(z)a^3(z)(\pi(z)-\varphi(z))^2\right)
 \ee
where $V(q,z)$ is the vector bulk-to-boundary propagator which satisfies
the following equation

\be\label{veom1}
  -\p_z^2V(q,z) +\omega'\p_zV(q,z) = q^2 V(q,z),
  \ee
with the boundary condition $V(q,z\to 0)=1$, so that $V(q,z)$ only
depends on $q^2$. The general considerations and derivations for the
electromagnetic form factor in the AdS space can be seen in
literature \cite{PS2,GR}, which may be compared with the
corresponding light-front form given in \cite{BT1,BT2,BT3}. Our
result is plotted in Fig.~\ref{fig:gfpi} which shows a good
agreement with the experimental data. It appears to be better than
the one obtained in the AdS/QCD models considered in
\cite{Gherghetta:2009ac} and \cite{Kwee:2007nq}. We would like to
point out that though the coupling constant $g_{\rho\pi\pi}$ in
Eq.\,(\ref{grhopipi}) and the form factor $F_\pi$ in
Eq.\,(\ref{Fpi}) have a similar expression, the latter agrees
well with the experimental data and the former has a discrepancy of
about a factor of 2 with the experimental result. The reason is
that the form factor is obtained by considering contributions from
all resonance mesons rather than only from the ground state $\rho$
meson, which may explicitly be seen from the following expression
\cite{Kwee:2007nq},
   \be\label{anotherFpi}
   F_\pi(q^2)=-\sum_{n=1}^{\infty}\frac{f_ng_{n\pi\pi}}{q^2-M_n^2}.
   \ee
where the summation is over the ground state $\rho$ meson and its
excited states, $f_n$ are related to their decay constants, $g_{n\pi\pi}$ are the corresponding
coupling constants and $M_n$ are their masses. It is noticed that the pion form
factor is not sensitive to the IR boundary conditions, but it is
relevant to the magnitude of quartic interaction. Thus more precise
experimental measurement may be used to determine the coupling
constant $\lambda$.

\section{Conclusion}

In this paper, we have shown how the chiral symmetry breaking and
linear confinement can be incorporated in the soft-wall AdS/QCD
model by simply modifying the 5D metric at the IR region. This is
realized because a modification of the 5D metric at the IR region
allows us to yield the desired IR and UV behaviors for the
background dilaton field, so that the linear trajectories for the
resonance meson states are simply obtained, and the resulting
resonance meson states agree remarkably with the experimentally
confirmed resonance states. It has also been shown that when a
physically reasonable form of the bulk VEV is constructed to satisfy
the boundary conditions of $AdS_5$, the resulting mass spectra for
all the resonance meson states are not very sensitive to the choice
of the exact forms of the bulk VEV, which can be seen from the
tables and figures of mass spectra by comparing among three models
I, II and III. While the dependence of mass spectra on the IR
boundary conditions of the bulk VEV has been shown to be significant
for the scalar and vector mesons, and sizable for the pseudoscalar
and axial-vector mesons, which can also be seen from the tables and
figures of mass spectra by comparing two cases in each model, i.e.,
between Ia and Ib, IIa and IIb, IIIa and IIIb, the cases $a$ and $b$ correspond
 to the IR boundary conditions of bulk VEV $v(z\to
\infty)\sim z$ and $v(z\to \infty)\sim \sqrt{z}$, respectively. It
is concluded that the case $b$ with the IR behavior $v(z\to
\infty)\sim \sqrt{z}$ appears to provide more reasonable predictions
for all the resonance mass spectra, which can be seen explicitly
from Figs.~\ref{fig:gmass1} and~\ref{fig:gmass2}.

The effects of the quartic interaction term have been investigated
in detail; a physically reasonable sign of the coupling constant
for the quartic interaction, which is opposite to the one considered
in \cite{Gherghetta:2009ac}, is taken to avoid the possible
instability of the bulk scalar potential. As a consequence, it has
been found that the introduction of the quartic interaction with an
appropriate sign and reasonable magnitude can result in a better
agreement for the resonance mass spectra of scalar, pseudoscalar,
vector and axial-vector mesons. Quantitatively, the agreement with
the experimental data is found to be within $10\%$ for all
resonance meson states, though the predictions for the ground state
mesons of scalar, vector and axial vector are not as good as the
ones for the resonance states.

As such a simply modified soft-wall AdS/QCD model can result in such
remarkable predictions for the mass spectra of resonance mesons, it
would be interesting to further study the dynamical origin of the
metric induced conformal symmetry breaking in the IR region. Also in
the present consideration, the dilaton and gravity are treated as
background fields; it would also be interesting to further
investigate the important role of the dilaton field in connection
with the stringy features of QCD and the 5D gravity effects from the
back-reacted geometry which has been studied in a class of hard-wall
AdS/QCD models~\cite{SWWX}. It is a necessity to study the possible
higher order interaction terms and their effects on the mass spectra
and form factors of various mesons. It is also natural to extend to
the three flavor case and consider the $SU(3)$ breaking and
instanton effects.

 % lookout the further develop of this field.

%%%%%%%%%%%%%%%%%%%%%%%%%%%%%%%%%%%%%%%%%%%%%%%%%%%%%%%%%%%%%%%%%%%%%%
\section*{Acknowledgements}

This work was supported in part by the National Science Foundation
of China (NSFC) under Grant \#No. 10821504 and the Project of
Knowledge Innovation Program (PKIP) of the Chinese Academy of Science.

%%%%%%%%%%%%%%%%%%%%%%%%%%%%%%%%%%%%%%%%%%%%%%%%%%%%%%%%%%%%%%%%%%

\newpage

\end{document}